
\input harvmac.tex
%
%
\def\sign{{\rm sign}}
\def\muN{{\mu_1 \dots \mu_N}}
%
%
\lref\dissqm{A.~O.~Caldeira and A.~J.~Leggett, Physica  {\bf 121A}(1983) 587;
Phys. Rev. Lett. {\bf 46} (1981) 211; Ann. of Phys. {\bf 149} (1983) 374.}
\lref\cgcdef{C.~G.~Callan and D.~Freed, Nucl. Phys {\bf B374}(1992)543.}
\lref\osdqm{C.G. Callan, L. Thorlacius,  Nucl. Phys. {\bf B329} (1990) 117.}
\lref\cff{C.~G.~Callan, A.~G.~Felce and D.~E.~Freed, Nucl. Phys. {\bf B392}
(1993) 551}
\lref\CTWI{C.~G.~Callan, L.~Thorlacius, Nucl.  Phys. {\bf B319} (1989) 133.}
\lref\clny{C.~G.~Callan, C.~Lovelace, C.~R.~Nappi, and S.~A.~Yost,
Nucl. Phys. {\bf B293} (1987) 83; Nucl. Phys. {\bf B308} (1988) 221.}
\lref\schmid{A.~Schmid, Phys. Rev. Lett. {\bf 51} (1983) 1506.}
\lref\ghm{F.~Guinea, V.~Hakim and A.~Muramatsu, Phys. Rev. Lett.
{\bf 54} (1985) 263.}
\lref\fisher{M.~P.~A.~Fisher and W.~Zwerger, Phys. Rev. {\bf B32} (1985) 6190.}
\lref\contact{D.~E.~Freed, Nucl. Phys. {\bf B409}(1993)565.}
\lref\wardid{D.~E.~Freed, Nucl.~Phys.~{\bf B424} (1994), 628.}
\lref\proceeding{D.~E.~Freed, to appear in ``Strings '93", M.~B.~Halpern,
G.~Rivlis, A.~Sevrin, eds., World Scientific, CTP \#2251, hep-th/9311034.}
\lref\cfw{C.~C.~Chamon, D.~E.~Freed, and X.-G.~Wen,
cond-mat/9408064, to appear in Phys.~Rev.~{\bf B}.}
\lref\kanefisher{C.~L.~Kane and Matthew P.~A.~Fisher,
Phys.~Rev.~Lett. {\bf 68},
(1992) 1220; Phys.~Rev.~B {\bf 46}, (1992) 15233.}
\lref\cardy{J.~L.~Cardy, Nucl.~Phys.~ {\bf B240} (1984) 514.}
\lref\ghoshal{S.~Ghosahl and A.~B.~Zamolodchikov, Int.~J.~Mod.~Phys. {bf A9}
(1994) 3841.}
\lref\affleck{I.~Affleck and A.~Ludwig, Nucl.~Phys.~ {\bf B352} (1991) 841.}
\lref\wen{X.-G.~Wen, Intl.~J.~of Mod.~Phys.~B {\bf 6}, (1992) 1711; and
references therein.}
\lref\acny{A.~Abouelsaood, C.~G.~Callan, C.~R.~Nappi and S.~A.~Yost,
Nucl.~Phys.~{\bf B280} (1987) 599.}
\lref\fradkin{E.~Fradkin and A.~Tseytlin, Phys.~Lett.~{163B} (1985) 123.}
\lref\cgcirk{C.~G.~Callan and I.~R.~Klebanov, Phys.~Rev.~Lett.~ {\bf 72}
(1994) 1968; C.~G.~Callan, I.~R.~Klebanov, A.~Ludwig and J.~M.~Maldacena,
Nucl.~Phys.~{\bf B422} (1994) 417.}
\lref\thorpol{J.~Polchinski and L.~Thorlacius, Phys.~Rev.~{\bf D50}
(1994) 622.}
\lref\ckmy{C.~G.~Callan, I.~R.~Klebanov, J.~M.~Maldacena, A.~Yegulalp,
hep-th/9503014.}

%
%
{\nopagenumbers

\baselineskip 12pt plus 1pt minus 1pt
\centerline{\bf  EXACT SOLUTIONS FOR CORRELATION FUNCTIONS}
\smallskip
\centerline{{\bf IN SOME 1+1 D FIELD THEORIES WITH BOUNDARY}\footnote{*}{This
work is supported in part by
funds provided by the U. S. Department of Energy (D.O.E.) under cooperative
agreement \#DE-AC02-76ER03069 and
by National Science Foundation grant PHY9218167.  D.~F.~is currently a
Bunting Fellow sponsored by the Office of Naval Research.}}
\vskip 24pt
\centerline{Denise E.~Freed}
\vskip 12pt
\centerline{\it Center for Theoretical Physics}
\centerline{\it Laboratory for Nuclear Science}
\centerline{\it and Department of Physics}
\centerline{\it Massachusetts Institute of Technology}
\centerline{\it Cambridge, Massachusetts\ \ 02139\ \ \ U.S.A.}
\vskip .7in
\centerline{\bf Abstract}
\smallskip
We consider 1+1 D theories which are free everywhere
except for cosine and magnetic interactions on the boundary.
These theories arise in dissipative quantum systems, open string theory,
and, in special cases, tunneling in quantum Hall systems.
These boundary systems satisfy an approximate SL(2,Z) symmetry as a
function of flux per unit cell and dissipation.
At special multicritical points,
they also satisfy a set of reparametrization Ward identities and
have homogeneous, piecewise-linear correlation functions in momentum space.
In this paper, we use these symmetries to find exact solutions for
some of the correlation functions.  We also comment on the form of
the correlation functions in general, and verify that the SL(2,Z)
duality transformation between different critical points is satisfied
exactly in all cases where the full solution is known.
\vskip .8in
\centerline{Submitted to: {\it Nuclear Physics B\/}}
\vfill
\vskip -12pt
\noindent CTP\#2422, HUTP-95/A011, hep-th/9503065\hfill March 1995
\eject}
\pageno=1
\baselineskip 18pt plus 2pt minus 1pt

\newsec{Introduction}
Recently, much attention has been paid to 1+1 dimensional field theories
with boundary \refs{\cardy, \ghoshal}.  Systems where the fields are free
everywhere except for an interaction at the boundary can be used to
describe dissipative quantum mechanics \refs{\dissqm, \osdqm},
tunneling of edge states in quantum Hall systems \refs{\wen,\kanefisher},
the Kondo problem \refs{\affleck}, and open string theory
with one boundary \refs{\clny, \acny, \fradkin}.
In this paper we will consider  the
particular case when there are two fields, and at the boundary each field
experiences a cosine potential and also a ``magnetic" potential that induces
an interaction between the two fields.  In the dissipative quantum system,
this describes a particle confined to two dimensions that is subject to a
doubly
periodic potential and a transverse magnetic field.  Restricting to only
one field (and setting the magnetic interaction to zero) this model is of
further interest because it describes the tunneling in quantum Hall
systems when the edge states are composed of only a single branch.
As indicated in ref.\cgcdef, we expect this system to have rich
behavior.  It has an approximate SL(2,Z) duality symmetry as a function
of magnetic flux per unit cell, $\beta$, and dissipation per unit cell,
$\alpha$.  At the
critical points, it should not only be scale invariant, but also satisfy a
set of Ward identities reflecting the symmetry under reparametrization of
the associated open string boundary state \CTWI.
The simplest of these just
requires the dissipative quantum system to be SL(2,R) invariant at its
critical points. In references \wardid, \cgcirk, and \ckmy\ it is shown that
at special values of flux and dissipation, these reparametrization Ward
identities are satisfied, but at many other proposed critical points
we do not yet  know that this is the case.  These ``special" values of flux
and friction all occur at multicritical points on the proposed phase
diagram in reference \cgcdef, and they have the property that the
magnetic interaction between exponentials of the two fields vanishes.

The only critical point occuring when the magnetic field is zero  is
described by a $c=1$ conformal field theory \refs{\cgcirk, \thorpol}.
It can be described by a system of free fermions \refs{\ghm, \cgcdef,
\thorpol}, but the simplest such representations
do not properly obey the duality symmetry combined with the SL(2,R)
invariance.  In ref. \contact, a more involved calculation, based on the
idea of the fermionization, shows that at all the special points,
the correlation functions in Fourier space must be homogeneous,
piecewise-linear funtions of the momentum.

In this paper we show how these various symmetries can be used to find exact
solutions for the correlation functions at the special points.  Because the
duality transformation relates these points to the $c=1$ conformal field
theory with zero magnetic field, we expect that even at non-zero magnetic
field the system should have a simple conformal field theory interpretation,
and it is hoped that the results of this paper will be a guide in finding
such theories.

Another issue we address in this paper is that, when the magnetic field is
equal to zero, all the connected correlation functions
other than the two-point function consist only of contact terms.
The duality transformation takes these correlation functions to ones with
non-zero magnetic field that are not contact terms.  This suggests that
the contact terms are physical and that the symmetries of the system will
suffice to fix them.  We have found that this is the case for many of
the correlation functions, and it seems likely it is true for all of them.
The final issue we address is to what extent the duality symmetries found in
ref. \cgcdef\ are exact.  We find that whenever we can solve exactly
for the correlation functions, the symmetry corresponding to $z \to z/(1+inz)$
is exact, where $z = \alpha + i\beta$.
In addition, we identify the value of the strength of the cosine potential
at which the system satisfies the conditions for
self-duality under the transformation $z \to 1/z$.

In Section 2, we describe the boundary systems studied in this paper,
and in Section 3 we review
all the symmetries and properties of the system that were
found in references \cgcdef, \contact, and \wardid.  Because we are solving
for contact terms, we find it more convenient to work in Fourier space,
where the contact terms are all well-defined functions.  Thus, in the
following two sections we first show how the SL(2,R) symmetries and
the Ward identities act on the system in Fourier space, and
then derive properties that correlation functions in momentum space
must have if they are to respect these symmetries.
Finally, in Section 6 we find the exact solutions for several of the
correlation functions and discuss the form of the general solution.
However, in the general case we still cannot prove that the symmetries are
enough to determine the solutions.

This paper largely explains the results given in reference \proceeding\
and gives the details of their derivations.  While completing this paper,
the author became aware of reference \ckmy, which was partly
motivated by this work.


\newsec{Background}
The dissipative Hofstadter model describes a quantum particle confined to
two dimensions subject to a doubly periodic potential, a perpendicular
magnetic field, and dissipation.  When the Caldeira-Leggett \dissqm\
model is used to model the dissipation, the Euclidean action for this system
is given by
\eqn\Sdef{S = S_q + S_\eta + S_V,}
where $S_q$ is the usual action of a particle in a constant magnetic field,
$S_\eta$ is a nonlocal kinetic term that gives the effect of the friction,
and $S_V$ is due to the periodic potential.  $S_q$ is given by
\eqn\Sqdef{S_q = \int_{-T/2}^{T/2} dt \left[{M\over 2} \dot{\vec x}^2
              +{ieB\over 2c} (\dot x y - \dot y x)\right].}
In this equation, $x(t)$ and $y(t)$ are the coordinates of the particle,
$B$ is the strength of the magnetic field, and $M$ is the mass of the
particle.  In the presence of the dissipation, $M$ just acts as a regulator,
so in the calculations of references \wardid\ and \contact\ we set $M$ to zero
and chose a more convenient regulator.  The part of the action due to the
periodic potential is chosen to be
\eqn\SVdef{S_V = \int_{-T/2}^{T/2}
      \left[V_0 \cos\left({2\pi x(t) \over a}\right)
           +V_0 \cos\left({2\pi y(t) \over a} \right)\right],}
where $V_0$ is the strength of the potential and $a$ is the size of the unit
cell.  The dissipative term in the action, due to the particle's interaction
with its environment, has the form
\eqn\Setadef{S_\eta
    = {\eta \over 4\pi} \int_{-T/2}^{T/2}\int_{-\infty}^\infty
    dt\, dt' \left({\vec x(t) -\vec x(t')\over t-t'}\right)^2.}
This term comes about by modeling the particle's environment with a bath of
harmonic oscillators which interact linearly with the particle.  The
functional integral over the oscillators induces the nonlocal term
\Setadef\ in the action for $\vec x(t)$.

Because we obtained this action by integrating over modes of free
oscillators, it is not surprising that this action can also be obtained from
a 1+1 dimensional system with a boundary, where the fields are free in the
bulk and interact only at the boundary \osdqm.  The action for
this 1+1 dimensional boundary theory is given by
\eqn\Sbulk{S_B = S_b + S_q + S_V,}
where
\eqn\Sbdef{S_b = {\alpha\over 8\pi} \int_{-T/2}^{T/2} \int_{-\infty}^0
              d\sigma dt \left((\partial_\mu x)^2 + (\partial_\mu y)^2 \right)}
is the bulk action;
\eqn\SqBdef{S_q = {i\pi\beta}\int_{-T/2}^{T/2} dt (\dot x y - \dot y x)}
is the boundary magnetic field term; and
\eqn\SVdef{S_V = V_0\int_{-T/2}^{T/2}
      \left[\cos\left(x(t)\right)
           + \cos\left(y(t)\right)\right]}
is the boundary action from the cosine potential.  In terms of the
parameters of the dissipative system,
$\alpha$ is the friction/unit cell, given by
\eqn\alphadef{2\pi\alpha = {\eta  a^2\over\hbar},}
$\beta$ is the magnetic flux per unit cell,  given by
\eqn\betadef{2\pi\beta  = {eB\over \hbar c}a^2,}
and $x(t)$ and $y(t)$ have been rescaled by $a/(2\pi)$.
When $\beta = 0$ and we let $\vec x(t) = x(t)$, then the action in equation
\Sbulk\ also describes the tunneling of edge states in the quantum Hall
effect, where $\alpha$ is related to the filling fraction.

In this paper, we are interested in calculating the correlation functions of
$\dot x(t)$ and $\dot y(t)$.  We will concentrate on the correlation
functions with $\dot  x$ and $\dot y$ located only on the boundary, because
these are the correlation functions of interest  in the dissipative quantum
system,  and because it is fairly straightforward to obtain the bulk
correlation functions once we can calculate  the boundary ones.  Thus, we
will be calculating functions of the form
\eqn\Cdef{\eqalign{C^{\mu_1\dots\mu_m}(t_1, \dots, t_m)
        =& \langle \dot x^{\mu_1}(t_1) \dots \dot x^{\mu_m}(t_m)\rangle \cr
        =& \int {\cal D} x(t)  \prod_{i=1}^m \dot x^{\mu_i}(t_i)
           e^{-{1\over\hbar}S}.}}
Because we expect many of the correlation functions to be contact terms, we
will find it more convenient to work in Fourier space, where the contact
terms are well-defined functions.  Thus we will solve for the correlation
functions of the form
\eqn\tCdef{\tilde C^{\mu_1\dots\mu_m}(k_1, \dots, k_m)
     = {1\over T^m} \int_{-T/2}^{T/2} \prod_{j=1}^m dt_j
       \langle \dot x^{\mu_1}(t_1) \dots \dot x^{\mu_m}(t_m)\rangle
       e^{-{2\pi i\over T} k_1t_1} \dots e^{-{2\pi i\over T}  k_m t_m}.}
Most of the calculations in references \contact\ and \wardid\ were done
for finite
values of $T$, so that $k_i$ take on only integer values.  However, for most
of our final answers, we will take the limit as $T \to \infty$.  Also, for
convenience, we will appropriately rescale the $\dot{\vec x}$'s and $V_0$
by $T$ so that the explicit dependence on $T$ drops out of all the
equations.

In references  \refs{\cff, \cgcdef, \contact, \wardid},
the cosine potential in equation \SVdef\ was treated perturbatively, so that
the correlation functions in \Cdef\ become integrals over correlation
functions with arbitrary numbers of the insertions of $e^{\pm i x(t)}$ and
$e^{\pm i y(t)}$.  These insertions of $e^{i q_x x(t)}$ and $e^{i q_y y(t)}$
behave like a Coulomb gas, with charges $q_x = \pm 1$ and $q_y = \pm 1$.
The $x$-charges interact logarithmically and the $y$-charges interact
logarithmically.  The only interaction between the $x$ and $y$ charges is
a phase, so that when the locations of the $x$ and $y$ charges are
interchanged, the correlation function picks up the phase
$\exp(\pm i 2\pi {\beta\over \alpha^2 + \beta^2} q_x q_y)$. Whenever
$\alpha / (\alpha^2 + \beta^2) = 1$ and $\beta/\alpha$ is an
integer, the dimension of the charge operators,
$e^{i q_x x(t)}$ and $e^{i q_y y(t)}$, is one and the phase is also
equal to one.  These special points are the multicritical points in the
phase diagram of reference \cgcdef\ that lie on the intersections
of the critical circle with $\alpha /(\alpha^2 + \beta^2) = 1$ and all
circles tangent to it.  In this paper, we will concentrate on these special
points.  For the special point with $\beta = 0$, the model
(with $\vec x(t) = x(t)$) is a $c=1$ conformal field theory
\refs{\cgcirk, \thorpol}.  In the picture of the tunneling of
edge states, this theory just corresponds to tunneling of free fermions.
Because the duality transformation relates the other special critical points
to the one at $\beta = 0$, and because the theory at these special
critical points satisfies a set of reparametrization invariance Ward
identities, we expect that these theories should also be fairly simple
conformal field theories.  However, unlike in the $\beta = 0$ case,
the connected correlation functions should not consist only of contact
terms \cff,  so the conformal field theories should not be
as trivial as when $\beta = 0$.

The correlation function should depend on both  the potential strength,
$V_0$, and the value of $\alpha$ and $\beta$.  Because each of the
special points has a different value of the flux $\beta$, we will label
the correlation functions by the value of $\beta$ at the point, and call
the correlation function
$\tilde C^{\mu_1\dots \mu_m}(k_1, \dots, k_m;\beta)$.
However, for convenience, we will omit the $V$ dependence in this notation.

In the following sections we will review and explain the results of
references \contact, \cgcdef, and \wardid\ about the properties of the
correlation functions.  Most of the properties were derived using the
Coulomb gas picture, and we refer the reader to those references for details
of their derivations.

\newsec{Symmetries of the System}
In this paper we will use the various symmetries and properties of this
system to find exact solutions for the correlation functions.  We will begin
by reviewing these symmetries which were derived in references \cgcdef,
\contact, and \wardid.  In the latter two references,
all the calculations were done perturbatively
to all orders in $\alpha'$ (where $\alpha' = 1/\alpha$),
and the results are valid at {\it every} order in $V$.

\subsec{Duality Symmetry}
In ref. \cgcdef, we show that this system has an approximate duality symmetry
under $z \to z + i$ and $z \to 1/z$, where $z = \alpha + i\beta$.  Under
these transformations, the coordinate correlation functions transform in
a simple way.  In particular, up to questions about renormalization, we
expect the symmetry under $z \to z/(1 + inz)$ to be exact.  In this case,
the two-point function at the special multicritical points with $z =
1/( 1 + n^2) + in/(1+n^2)$ and potential strength $V_0$ can be obtained
from the two-point function at zero magnetic field with $z = 1$ and
potential strength $V_0$ by the following transformation \cgcdef:
\eqn\xxbxxz{\eqalign
{\langle \dot{\tilde x}^\mu(k) \dot{\tilde x}^\nu(-k) \rangle (z, V_0)
    =& |k| \left[ ({\beta\over\alpha})^2 \delta^{\mu\nu}
         + {\beta\over\alpha} \epsilon^{\mu\nu}\sign(k)\right]\cr
     &+ r^{\mu\rho}(k)r^{\nu\sigma}(-k)
        \langle \dot{\tilde x}^\rho(k) \dot{\tilde x}^\sigma(-k) \rangle
           (1, V_0),}}
where
\eqn\rmndef{r^{\mu\nu}(k) = \delta^{\mu\nu} - {\beta\over\alpha}
           \sign(k) \epsilon^{\mu\nu}.}
Because the value of the potential remains the same, and because, for
the multicritical points we are considering, $z$ is entirely determined
by its imaginary part $\beta$, we will drop the argument $V_0$ and
replace the argument $z$ by $\beta$.  Also we will use the notation
described in the previous section,
$\tilde C^{\mu\nu}(k_1, k_2;\beta)$, for the two-point function.
Then, because the two-point function has no off-diagonal terms when
$\beta = 0$, we can write equation \xxbxxz\ as
\eqn\Ctptrans{\tilde C^{\mu\nu}(k, -k; \beta) =
       \left[\prod_{j=1}^2 r^{\mu_jx}(k_j)
      + \prod_{j=1}^2 r^{\mu_jy}(k_j)\right]\tilde C(k, -k; 0)
      + |k| \left[ ({\beta\over\alpha})^2 \delta^{\mu\nu}
         + {\beta\over\alpha} \epsilon^{\mu\nu}\sign(k)\right].}
Similarly, for higher m-point functions, we expect that \contact
\eqn\Cnptrans{\tilde C^{\mu_1\dots\mu_m}(k_1, \dots k_m; \beta)
   = \left[\prod_{j=1}^m r^{\mu_j x}(k_j)
      + \prod_{j=1}^m r^{\mu_j y}(k_j)\right]
         \tilde C(k_1,\dots k_m; 0).}
(A similar transformation should hold for correlation functions at any
$z$ and $z'$ that are related by $z' = z/(1+inz)$.)
Once we are careful about how the theory is regulated, the $\tilde C(k_1,
\dots, k_m;0)$ in these expressions must be replaced with
\eqn\CtoF{\tilde C(k_1, \dots, k_m;0) \to F(\vec k; \beta).}
Note that $F$ depends on the size of the magnetic field, $\beta$, but it
is independent of the indices $\mu_1, \dots, \mu_m$.  For the original
duality symmetry to be exact, $F(\vec k; \beta)$ must be {\it independent}
of $\beta$.  We will see that when the symmetries are enough to determine
the correlation functions, then the solution for $F(\vec k; \beta)$
is unique up to a constant, regardless of the value of $\beta$.

The duality transformation under $z \to 1/z$ for an arbitrary $m$-point
function can be found using the methods of reference \cgcdef.
Unlike the transformation for $z \to 1/(1+inz)$, when $z$ goes to $1/z$
the value of $V_0$ also changes.  The
self-dual point under this transformation occurs at $z=1$ and some
particular value of $V_0$ that will be determined in Section 6.4.
At the self-dual point, the correlation functions satisfy the following
relation:
\eqn\sdual{\left(1-\prod_{i=1}^m\sign(k_i)\right)
      \tilde C(k_1, \dots, k_m;0) = 0.}
This means that at the self-dual point we expect $\tilde C(\vec k)$ to
vanish if an odd number of the $k_i$ are equal to zero.   Although
the derivation for the transformation under $z \to 1/z$ involved many
approximations, we will see that at the ``self-dual" value of $V_0$,
equation \sdual\ is satisfied exactly.

\subsec{Piecewise Linearity}
At the special multicritical points, the unregulated partition function can
be written as one for a fermion gas with quadratic interactions, which leads
us to expect the theory is solvable.  Furthermore, when $\beta=0$, the
expressions for the correlation functions can also be written in terms of
bilinears of the fermions, and we can obtain exact solutions for all the
correlation functions of the $\dot x$'s and $e^{ix(t)}$'s.\footnote{*}
{The subtleties of the fermionization of this system
have also been considered in reference \thorpol.}  The problem
with these solutions is that the fermionization is valid only for large
distance behavior; it does not necessarily give the correct short-distance
behavior.  This becomes a problem once we are trying to solve for
correlation functions that contain contact terms.  Also, when $\beta\ne0$,
we run into difficulties when considering correlation functions such as
$\langle \dot x(t) y(0)\rangle$, because these are just
delta-functions of $t$, and therefore only have short-distance behavior.  To
overcome these difficulties and find correlation functions that satisfy both
the duality symmetries and the reparametrization invariance Ward identities,
in reference \contact\ we started with the regulated version of the theory.
Making use of the fact that the unregulated theory can be fermionized, we
found that the coordinate correlation functions in the regulated theory are
piecewise-linear, homogeneous functions of the momenta.  In particular,
\eqn\FaRk{F(\vec k; \beta) = \vec a_R(\vec k) \cdot \vec k,}
where $F(\vec k; \beta)$ is defined in the previous section
by equations \Cnptrans\ and  \CtoF, and $\vec a_R(\vec k)$
depends only on the signs of the sums
\eqn\sumk{\sum_{i\in S} k_i \quad {\rm with} \quad S \subset \{1, \dots, m\}.}
Also, from the calculations done in reference \contact, we can show that,
as a function over the reals, $\tilde C(\vec k)$, and therefore
$F(\vec k; \beta)$, is continuous.  It follows from equation \sumk\
that the slope, $\vec a_R (\vec k)$, changes only
when one of the hyperplanes, given by
\eqn\sumkzero
{\sum_{i\in S} k_i=0 \quad {\rm for} \quad S \subset \{1, \dots, m\},}
is crossed.

Because the functions are piecewise linear in momentum space, in real space
we expect them only to include $\delta$-functions and factors of $1/(t_i -
t_j)$.  If a function is homogeneous in momentum space, then it has an
additional reparametrization symmetry in real space.  In particular, it
tells us what happens under the transformation $z \to z^n$, where
$z = e^{2\pi i t/T}$.  We can derive
this symmetry as follows.  The linearity and homogeneity of the correlation
functions in momentum space means that they satisfy the following relation:
\eqn\linea{n \tilde C(k_1, \dots , k_n) = \tilde C(nk_1, \dots nk_n)
\quad {\rm for}\quad n \in Z^+.}
(For simplicity, in this section we are dropping the indices on $C$.)
Next, we can consider the following Fourier series
\eqn\Cftmess{\sum_{0\le j_l < n} C(z_1 e^{2\pi i j_1\over n}, \dots ,
                  z_m  e^{2\pi i j_m\over n})
               = \sum_{0\le j_l < n} \sum_{k_i = -\infty}^\infty
                 \tilde C(k_1, \dots , k_m) \prod_{l=1}^m
                  (z_l e^{2\pi i j_l\over n})^{k_l} .}
Interchanging the order of sums and products, and using the relation
\eqn\sumnZ{\sum_{0\le j < n}(e^{2\pi i k\over n})^j
               = \cases{n & for $ k \in nZ$ \cr
                        0 & for $k \notin nZ$,\cr}}
the expression on the right-hand side of \Cftmess\ becomes
\eqn\Cftmessr{n^m \sum_{k_i\in nZ} \tilde C(k_1, \dots, k_m) z_1^{k_1}
\dots z_m^{k_m}.}
In this expression, we can replace the sum over $k_i$ by a sum over
$l_i = k_i/n$ and apply equation \linea\ to obtain
\eqn\Cftmesss{n^{m+1} \sum_{l_i \in Z} \tilde C(l_1, \dots, l_m)
              (z_1^n)^{l_1} \dots (z_m^n)^{l_m}.}
Performing the sum, and setting the result equal to the left-hand side of
equation \Cftmess, we find
\eqn\zntrans{C(z_1^n, z_2^n, \dots, z_m^n) = {1 \over n^{m+1}}
               \sum_{0\le j_l < n} C(z_1 e^{2\pi i j_1\over n}, \dots ,
                 z_m e^{2\pi i j_m\over n}).}
Because each $z_l e^{2\pi i j_l/n}$ is an $n$th root of $z_l^n$, equation
\zntrans\ gives the transformation of the real-space correlation functions
when $z \to z^n$.

\subsec{Permutation and Inversion Symmetries and the Boundary Conditions}
In the derivation of equations \Cnptrans\ and \CtoF\ for $\tilde C$ in
terms of $F(\vec k; \beta)$,
several other symmetries for $F$ also followed.  The first is that
$F(\vec k; \beta)$ is symmetric under interchanges of $k_i$ and $k_j$.
The second is that $F(\vec k; \beta) = F(-\vec k; \beta)$.  Also,
whenever $\vec k$ has an odd number of components, we have the condition
that $F(\vec k; \beta) = 0$.

There is an additional property of $\tilde C$ and $F$ that was found
in reference \contact.  Both $\tilde C(\vec k)$ and $F(\vec k)$
must equal zero when any one of the $k_i = 0.$  As we shall see in Section
5.4, the SL(2,R) invariance requires $\tilde C(\vec k;\beta)$
to be continuous, for any value of $\beta$.
However, according to the duality transformation, $\tilde C(\vec k;\beta)$
is obtained from $\tilde C(\vec k; 0)$ by
multiplying it by factors of $\sign(k_i)$.
Therefore the boundary conditions at $k_i = 0$
turn out to be a necessary condition for the duality transformation
to be satisfied along with SL(2,R) invariance.

\subsec{The Reparametrization Invariance Ward Identities}
The 1-D field theory describing dissipative quantum mechanics also describes
a 2-D statistical theory with boundary.  At the critical points, we expect
the theory to be conformally invariant, and also to describe the boundary
state in open string theory \osdqm.  To be a conformal field theory,
the theory must satisfy a set of reparametrization invariance
Ward identities \CTWI.  These reflect
the fact that the boundary state must be invariant under reparametrizations
of the boundary.  Thus, the boundary state $|B\rangle$ must satisfy
\eqn\LtLB{\left(L_n - \tilde L_{-n}\right) |B\rangle = 0 \qquad {\rm for}
             \quad -\infty \le n \le \infty,}
where the $L_n$ and $\tilde L_{-n}$ are the closed string Virasoro generators,
given by
\eqn\Lndef{L_n = {1\over2} \sum_{m = -\infty}^\infty
                  \vec \alpha_{n-m}\cdot \vec \alpha_m,}
and similarly for $\tilde L_{-n}$.  For $-m < 0$, $\vec \alpha_{-m}$ and
$\tilde{\vec\alpha}_{-m}$ are the closed string creation operators. For
$m \ge 0$, $\vec\alpha_m$ can be expressed as a derivative with respect to
$\vec\alpha_{-m}$ as follows:
\eqn\alphamdef{\alpha_m^\mu = m {\partial\over \partial\alpha_{-m}^\mu},}
and similarly for $\tilde\alpha_m$.  The boundary state is given by
\eqn\Bdef{|B\rangle = \exp\left(\sum_{m=1}^\infty {1\over m}
                        \vec\alpha_{-m}\cdot\tilde{\vec\alpha}_{-m}
                     -W[\vec\alpha, \tilde{\vec\alpha}, \vec x_0]\right)
                     |0\rangle ,}
where $|0\rangle$ is the closed string vacuum state, $\vec x_0$ is the zero
mode of $\vec x(t)$, and
$W[\vec\alpha, \tilde{\vec\alpha}, \vec x_0]$ is the connected generating
functional for the 1-D system, given by
\eqn\WXzdef{\exp\left(-W[\vec\alpha, \tilde{\vec\alpha}, \vec x_0] \right)
         = \int\left[{\cal D} \vec x(s)\right]'\exp(-S_\eta-S_q-S_V-S_{LS}).}
In this equation, the prime denotes that the integration over the zero mode
of $x(t)$ is omitted.
$S_\eta$, $S_q$ and $S_V$ are defined in Section 2, and
$S_{LS}$ is the linear source term given by
\eqn\SLSalpha{S_{LS} = \sqrt{2\over\alpha'}\int ds\vec\alpha(s)\cdot
                                         \vec x(s),}
with
\eqn\alphamudef{\alpha^\mu(s)= \sum_{m=1}^\infty
       i\left(\tilde\alpha^\mu_{-m} e^{-ims} + \alpha^\mu_{-m}e^{ims}\right).}
The $\alpha'$ in equation \SLSalpha\ is the string tension.
Once we rescale $x(t)$ as we did in
equations \Sbdef-\SVdef, it is given by $\alpha' = 1/\alpha$.
By applying equation \LtLB\ to equation
\Bdef, Callan and Thorlacius showed \CTWI\ that the condition for
reparametrization invariance translates to a condition on only the
1-D theory on the boundary, $W[\vec\alpha, \tilde{\vec\alpha}, \vec x_0]$.
In reference \wardid, we show that this identity is
satisfied to all orders in $V$
for the cosine potential, as long as $\alpha/(\alpha^2+\beta^2) = 1$ and
$\beta/\alpha \in Z$.

In the dissipative quantum system, we instead integrate over the zero mode,
so that $W$ depends only on $\alpha_m$ and $\tilde \alpha_m$, and we have
$\partial W/ \partial x_0^\mu = 0$.  We will find it convenient to define
a source $J(t)$ that is coupled to $\dot x(t)$ by
\eqn\Jdef{\eqalign{
         J^\mu_m =&\ \ \sqrt{2\over\alpha'}{1\over m}\alpha^\mu_{-m}\cr
         J^\mu_{-m} =&-\sqrt{2\over\alpha'}{1\over m}\tilde\alpha^\mu_{-m},}}
for $m>0$. Then the generating function becomes
\eqn\WJdef{e^{W[J]} = \int{\cal D}\vec x(t) e^{-(S_\eta+S_q+S_V)}
                      e^{\int \vec J\cdot \dot{\vec x} dt},}
and the connected correlation functions are given by
\eqn\CJWdef{\tilde C^{\mu_1, \dots , \mu_l}(m_1, \dots, m_l)
     = \prod_{i=1}^l {\partial \over \partial J^{\mu_i}_{-m_l}} e^{W[J]}
        \bigg |_{J=0}.}
The Ward identity then reduces to
\eqn\Wardid{{2\over\alpha'}\sum_{m=1}^{n-1}\left[
      {1\over2}{\partial W\over \partial \vec J_{m}}\cdot
      {\partial W\over \partial \vec J_{-m+n}}
      - {\partial^2W\over\partial\vec J_m \cdot \partial \vec J_{-m+n}}\right]
     + \sum_{m=-\infty}^\infty m
       \vec J_{-m} \cdot {\partial W \over \partial \vec J_{-m+n}} = 0.}
and the results from reference \wardid\ showing that the multicritical points
satisfy these Ward identities still hold.  The last term in equation
\Wardid\ is precisely the change in $W$ under reparametrizations, and the
first two terms come from the fact that the $\eta$ term in the original
1-D action is not reparametrization invariant.

\newsec{Difference Equations from the Ward Identities}
In this section we will use the reparametrization invariance Ward identities
\Wardid\ to derive difference equations for the correlation functions in
momentum space.
\subsec{SL(2,R) invariance}
When $n = 0$, $\pm 1$, the reparametrizations in equation \LtLB\ are
simply the SL(2,R)
transformations.  (Strictly speaking, when $T$ is finite and the $k_i$ take
on values in the integers, these reparametrizations are the SU(1,1)
transformations.  However, in the limit as $T \to \infty$ they become
the SL(2,R) transformations instead.)
For these three values of $n$, the terms in the Ward identity
due to the breaking of reparametrization invariance are identically zero.
Thus the theory should be SL(2,R) invariant, and we expect that under these
transformations the $\dot x(t)'s$ within the correlation functions should
transform as dimension one operators.
In this section, we will see what conditions the SL(2,R) invariance of the
theory imposes on the correlation functions in momentum space.

The $n=0$ case of the Ward identity is
\eqn\nzWI{\sum_{m=-\infty\atop m\ne0}^\infty
              m J^x_{-m} {\partial W \over \partial J^x_{-m}}
       + \sum_{m=-\infty\atop m\ne0}^\infty
              m J^y_{-m} {\partial W \over \partial J^y_{-m}}
       = 0.}
To obtain the correlation functions, we take $N$ derivatives and then set the
$J$'s to zero.  The derivatives are
\eqn\derdef{\prod_{i=1}^N{\partial\over \partial J_{-k_i}^{\mu_i}}.}
Acting on the expression in equation \nzWI, they give
\eqn\nzWIcalc{\sum_{i=1}^N k_i {\partial^N W \over
                 \prod_{j=1}^N \partial J_{-k_j}^{\mu_j}} = 0,}
which can be written in terms of $\tilde C$ as
\eqn\nzCcalc{\sum_{i=1}^N k_i \tilde C^{\mu_1 \dots  \mu_N}(k_1,\dots,k_N)
              = 0.}
This implies that
\eqn\nzWIC{\tilde C^{\mu_1 \dots \mu_N}(k_1,\dots,k_N) = 0
     \quad {\rm unless} \quad \sum_{i=1}^N k_N = 0.}
This condition is the statement of conservation of momentum. It is just
what we expect from the $n=0$ reparametrization identity,
which tells us that the system is translationally invariant.

For the $n=1$ case, the Ward identity becomes
\eqn\noWI{\sum_{m = -\infty\atop m\ne0}^\infty  m \left[
               J_{-m}^x {\partial W\over\partial J_{1-m}^x}
             + J_{-m}^y {\partial W\over\partial J_{1-m}^y} \right] = 0.}
Again we take the derivatives in equation \derdef\ to obtain
\eqn\noWIcalc{\sum_{i=1}^N k_i
              {\partial W \over \prod_{j=1\atop j\ne i}^N
               \partial J_{-k_j}^{\mu_j} \partial J_{1-k_i}^{\mu_i}} = 0.}
Setting the $J$'s to zero, we find that
\eqn\noCWI{\sum_{i=1}^N k_i \tilde C^{\mu_1 \dots \mu_N}(\vec k-\hat e_i) = 0,}
where $\vec k = (k_1, \dots, k_N)$ and $\hat e_i$ is the unit vector
with a 1 in the $i$th component and zeroes everywhere else.  According to
the $n=0$ equation, in this equation the $k_i$'s must satisfy
$\sum k_i = 1$.

The $n=-1$ Ward identity similarly gives
\eqn\nmoCWI{\sum_{i=1}^N k_i \tilde C^{\mu_1 \dots \mu_N}(\vec k+\hat e_i) =
0.}

In solving for the correlation functions, it will be much simpler to write
everything in terms of the function $F(\vec k;\beta)$.  For convenience,
we will drop the $\beta$ in the argument of $F$.  According to the
definition of $F(\vec k)$, we have
\eqn\CRFdef{\tilde C^\muN(\vec k) = R^\muN(\vec k) F(\vec k)}
for $N>2$ and
\eqn\CRFtp{\tilde C^{\mu_1 \mu_2}(\vec k) = R^{\mu_1 \mu_2}(\vec k)F(\vec k)
          +|k_1| \left[({\beta\over\alpha})^2\delta^{\mu_1 \mu_2}
                 + {\beta\over\alpha} \epsilon^{\mu_1 \mu_2}\sign(k_1)\right]
                     \delta_{k_1,k_2}}
for $N=2$.  In these equations, $R$ is defined as
\eqn\Rdef{R^\muN = \prod_{i=1}^N r^{\mu_i x} + \prod_{i=1}^N r^{\mu_i y},}
where $r^{\mu\nu}$ is defined in equation \rmndef.  Also, it is important
to note that $R$ is never equal to zero and to recall that $F(\vec k) = 0$
if any of the $k_i = 0$.

The $n=0$ Ward identity in equation \nzWIC\ then implies that
\eqn\noFWI{\quad F(\vec k) = 0 \quad {\rm unless} \quad \sum k_i = 0.}
The $n=1$ Ward identity in equation \noCWI\ says that
\eqn\noRFWI{\sum_{i=1}^N k_i R^\muN(\vec k -\hat e_i) F(\vec k-\hat e_i) = 0.}
(This equation is satisfied even when $N=2$, because in that case the second
term of the correlation function \CRFtp\ satisfies the $n = 1$ Ward identity
by itself.)  According to the definition for $R$, we have
\eqn\RkeRk{R^\muN(\vec k - \hat e_i) = R^\muN(\vec k),}
as long as $k_i \ne 1$ and $k_i \ne 0$.  When $k_i = 1$ or $0$, we have
$k_i F(\vec k - \hat e_i) = 0$, so even in this case we can replace
$R^\muN(\vec k -\hat e_i)$ with $R^\muN(\vec k)$ in equation \noRFWI.
With this substitution, equation \noRFWI\ becomes
\eqn\noRFcalc{R^\muN(\vec k) \sum_{i=1}^N k_i F(\vec k -\hat e_i) = 0.}
Because $R$ is non-zero, we find that, for all $\vec k$ with
$\sum_{i=1}^N k_i=1$,
\eqn\noFWIp{\sum_{i=1}^N k_i F(\vec k - \hat e_i) = 0.}

To summarize, in Fourier space the SL(2,R) symmetry implies that
\item{i)}  $F(\vec k) = 0$ unless $\sum_{i=1}^N k_i = 0.$
\item{ii)} $\sum_{i=1}^N k_i F(\vec k - \hat e_i) = 0,$ \hfill \break
\noindent{and, similarly,}
\item{iii)} $\sum_{i=1}^N k_i F(\vec k + \hat e_i) = 0.$

\subsec{The Remaining Ward Identities}
Now we will turn our attention to the Ward identities for $|n| > 1$.  In
general, after taking $N$ derivatives, the last term in the Ward identity
is
\eqn\WIlt{\sum_{i=1}^N k_i
              {\partial W \over \prod_{j=1\atop j\ne i}^N
               \partial J_{-k_j}^{\mu_j} \partial J_{n-k_i}^{\mu_i}} = 0.}
When we set $J$ equal to zero, this becomes
\eqn\WIClt{\sum_{i=1}^N k_i \tilde C^\muN (\vec k-n\hat e_i) = 0.}
If we take $N$ derivatives of the second term, we obtain
\eqn\WIst{-{2\over\alpha'} \sum_{m=1}^{n-1} \left[ {\partial^{N+2}W \over
    \prod_{i=1}^N \partial J_{-k_i}^{\mu_i} \partial J_{m}^x J_{n-m}^x}
   + {\partial^{N+2}W \over
    \prod_{i=1}^N \partial J_{-k_i}^{\mu_i} \partial J_{m}^y J_{n-m}^y}
    \right].}
When $J$ goes to zero, this reduces to
\eqn\WICst{-{2\over\alpha'} \sum_{m=1}^{n-1} \left[
         \tilde C^{\muN xx}(k_1, \dots,k_N, -m, m-n) +
         \tilde C^{\muN yy}(k_1, \dots,k_N, -m, m-n)\right].}
The first term of the Ward identity is
\eqn\WIftnd{{1\over\alpha'} \sum_{m=1}^{n-1} \left[
     {\partial W\over\partial J_{m}^x} {\partial W\over\partial J_{n-m}^x}
  +  {\partial W\over\partial J_{m}^y} {\partial W\over\partial J_{n-m}^y}
     \right].}
After taking $N$ derivatives and setting $J$ to zero, we will obtain a product
of two correlation functions, each with $N$ or fewer vertices.  If we let
$S = \{1,2,\dots, N\}$ and let $s$ run through all the subsets of $S$, then,
after taking $N$ derivatives, we can write the first term as
\eqn\WIft{{1\over\alpha'} \sum_{m=1}^{n-1} \sum_{s \subset S}
     \left[
     {\partial^{|s|+1} W\over\prod_{i\in s}\partial J_{-m_i}^{\mu_i}
       \partial J_{m}^x}
     {\partial^{N-|s|+1} W\over\prod_{i\notin s}\partial J_{-m_i}^{\mu_i}
       \partial J_{n-m}^x} + \quad {\rm same} \quad {\rm for} \quad y\right],}
where $|s|$ is the number of elements in $s$.  When we set the $J$'s to
zero, we obtain
\eqn\WICft{{1\over\alpha'}\sum_{m=1}^{n-1}\sum_{s\in S} \left[
        \tilde C_s^x(\vec k_s, -m) \tilde C_{\bar s}^x(\vec k_{\bar s}, m-n)
      + \tilde C_s^y(\vec k_s, -m) \tilde C_{\bar s}^y(\vec k_{\bar s}, m-n)
      \right],}
where $\vec k_s$ contains all the $k_i$ with $i\in S$, and $\vec k_{\bar s}$
contains all the remaining $k_i$.  In equation \WICft, $\tilde C_s^x$ stands
for $\tilde C^{\mu_{j_1} \dots \mu_{j_{|s|}} x}$ with $j_i \in s$, and
$\tilde C_{\bar s}^x$ has all the remaining indices.  $\tilde C_s^y$ and
$\tilde C_{\bar s}^y$ are defined similarly.

The full Ward identity is then
\eqn\WIC{\eqalign{\sum_{m=1}^{n-1} \big[
         \tilde C^{\muN xx}&(\vec k, -m, m-n) +
         \tilde C^{\muN yy}(\vec k, -m, m-n)\big]\cr
   = &  {1\over 2} \sum_{m=1}^{n-1}\sum_{s\in S} \left[
        \tilde C_s^x(\vec k_s, -m) \tilde C_{\bar s}^x(\vec k_{\bar s}, m-n)
      + \tilde C_s^y(\vec k_s, -m) \tilde C_{\bar s}^y(\vec k_{\bar s}, m-n)
      \right]\cr
   &  -{\alpha'\over2}\sum_{i=1}^N k_i \tilde C^\muN (\vec k-n\hat e_i).}}
This equation gives an expression for the $(N+2)$-point functions in terms of
$N$-point functions and products of two correlation functions, each with less
than $N+2$ arguments.  However, the left-hand side contains a {\it sum} of
$(N+2)$-point functions.  Because the sums with different values of $n$ are not
all linearly independent, we cannot use this equation to directly calculate
all of the correlation functions of $N+2$ variables in terms of correlation
functions with fewer variables.  However, at the very least, it will give us
the proper normalization of the higher correlation functions.

Again, we would like to write these equations in terms of the simpler function
$F(\vec k; \beta)$, which is independent of all the indices.  The expression
on the left-hand side of the Ward identity can be written in terms of $F$ as
\eqn\WIFRft{\sum_{m=1}^{n-1}\left[R^{\muN xx}(\vec k, -m, m-n)
         + R^{\muN yy}(\vec k, -m, m-n)\right] F(\vec k, -m, m-n; \beta).}
Using the definition of $R$ and $r^{\mu\nu}$ and the relations $-m < 0$
and $m-n <0$, we can simplify this expression to
\eqn\WIFft{\left(1 + ({\beta\over\alpha})^2\right)R^\muN(\vec k)
          \sum_{m=1}^{n-1}F(\vec k, -m, m-n; \beta).}

The first term on the right-hand side of equation \WIC\ can be written in
terms of $F$ as follows:
\eqn\WIFRrsft{\eqalign{{1\over 2}\sum_{m=1}^{n-1}\sum_{s\subset S}\bigg\{
 & \left[\prod_{i\in s} r^{\mu_i x}(k_i)r^{xx}(-m)
         + \prod_{i\in s} r^{\mu_i y}(k_i)r^{xy}(-m)\right]\cr
 & \times \left[\prod_{i\notin s} r^{\mu_i x}(k_i)r^{xx}(m-n)
         + \prod_{i\notin s} r^{\mu_i y}(k_i)r^{xy}(m-n)\right]\cr
 &  \times F(\vec k_s, -m; \beta) F(\vec k_{\bar s}, m-n; \beta)\cr
  + & {\rm same} \quad {\rm for} \quad y \bigg\}.}}
To obtain this expression, we assumed that neither $F(\vec k_s, -m;\beta)$
nor $F(\vec k_{\bar s}, m-n;\beta)$ are two-point functions.  They can
be two-point functions only when $0 < k_i <n$ for some $k_i$, and we will
return to this special case shortly.  Again we can use the relations
$-m <0$ and $n-m<0$ and the definition for $r^{\mu\nu}$ to simplify this
expression.  We obtain
\eqn\WIFrsft{\left(1 + ({\beta\over\alpha})^2\right){1\over2}
         R^\muN(\vec k) \sum_{m=1}^{n-1}\sum_{s\subset S}
         F(\vec k_s, -m; \beta) F(\vec k_{\bar s}, m-n; \beta).}
The last term of the Ward identity can be written in terms of $F$ as follows:
\eqn\WIFRlt{-{\alpha'\over 2}\sum_{i=1}^N k_i R^{\muN}(\vec k -n\hat e_i)
              F(\vec k - n\hat e_i;\beta).}
As long as $k_i \le 0$ or $k_i \ge n$, we can replace
$R^\muN(\vec k-n\hat e_i)$ with $R^\muN(\vec k)$ to obtain
\eqn\WIFlt{-{1\over 2}\left(1 + ({\beta\over\alpha})^2\right)
         R^\muN(\vec k)\sum_{i=1}^N k_i
              F(\vec k - n\hat e_i;\beta).}
In this equation, we have made use of the identities $\alpha' = 1/\alpha$
and, on the critical circle, $\alpha/(\alpha^2 + \beta^2) = 1$.

Therefore, we find that as long as none of the $k_i$ have values $0<k_i<n$,
the Ward identity is
\eqn\WIRF{\eqalign{\left(1 + ({\beta\over\alpha})^2\right)&R^\muN(\vec k)
          \sum_{m=1}^{n-1}F(\vec k, -m, m-n; \beta) \cr
     & = \left(1 + ({\beta\over\alpha})^2\right){1\over2} R^\muN(\vec k) \cr
   &\,\, \times    \left[\sum_{m=1}^{n-1}\sum_{s\subset S}
         F(\vec k_s, -m; \beta) F(\vec k_{\bar s}, m-n; \beta)
    -   \sum_{i=1}^N  F(\vec k - n\hat e_i;\beta)\right].}}
Because $R$ is non-zero, the Ward identity reduces to
\eqn\WIF{\sum_{m=1}^{n-1}F(\vec k, -m, m-n; \beta)
         = {1\over 2} \sum_{m=1}^{n-1} \sum_{s\subset S} F(\vec k_s, -m;\beta)
               F(\vec k_{\bar s}, m-n; \beta)
          -{1\over 2} \sum_{i=1}^N k_i F(\vec k -n\hat e_i;\beta).}

To finish, we must show that this equation remains true even when $0 < k_l < n$
for any $k_l$.  In that case, the two-point functions $C^{\mu_l \nu}(k_l, -m)$
and $C^{\mu_l \nu}(k_l, m-n)$ are non-zero and have the form given in equation
\CRFtp.  Accordingly, they have a piece of the form we assumed in deriving
equation \WIF, but also an additional piece, of the form
\eqn\tpcor{\left(({\beta\over\alpha})^2\delta^{\mu_l\nu}
           +{\beta\over\alpha}\sign(k_l)\epsilon^{\mu_l\nu}\right)
            |k_l|.}
This additional piece will give a correction to the second term in the
Ward identity whenever $0<k_l < n$, which has the form
\eqn\WICstcor{|k_l|\left(({\beta\over\alpha})^2\delta^{\mu_l\nu}
           +{\beta\over\alpha}\sign(k_l)\epsilon^{\mu_l\nu}\right)
           \tilde C_s^\nu(\vec k_s, k_l-n),}
where, in this case, $s = \{1,\dots, l-1,l+1,\dots,N\}$.  We can substitute
the expression
\eqn\CsrF{\tilde C_s^\nu(\vec k_s, k_l-n)
          = \left[r^{\nu x}(k_l-n) \prod_{i=1\atop i\ne l}^N r^{\mu_i x}(k_i)
               + r^{\nu y}(k_l-n) \prod_{i=1\atop i\ne l}^N r^{\mu_i y}(k_i)
            \right] F(\vec k -n\hat e_l)}
for $C$ in equation \WICstcor.  Then, if we simplify, using the definition
of $r^{\mu\nu}$ and the relations $k_l -n < 0$ and $k_l > 0$, we find the
following additional term in the Ward identity:
\eqn\WIFstcor{{\beta\over\alpha}\left(1+ ({\beta\over\alpha})^2\right)
               k_l F(\vec k - n\hat e_l)
          \left[\epsilon^{\mu_l x}\prod_{i=1\atop i\ne l}^N r^{\mu_i x}(k_i)
            +\epsilon^{\mu_l y}\prod_{i=1\atop i\ne l}^N r^{\mu_i y}(k_i)
           \right].}

We have also neglected what happens to the last term of the Ward identity
when $0<k_l < n$, for the $l$th term in the sum.  We can write the contribution
to the Ward identity from this term as follows:
\eqn\CFF{-{\alpha'\over 2} k_l C^\muN (\vec k -n\hat e_l)
               = {\cal F}_1 + {\cal F}_2,}
where
\eqn\Fonedef{{\cal F}_1
        = -{\alpha'\over 2} k_l R^\muN(\vec k) F(\vec k-n\hat e_l),}
and
\eqn\Ftwodef{{\cal F}_2 = -{\alpha'\over 2} k_l
               \left[ R^\muN(\vec k-n\hat e_l) - R^\muN(\vec k)\right]
               F(\vec k-n\hat e_l).}
${\cal F}_1$ has the same form as the contribution when $k_l \le 0$ or
$k_l \ge n$, so it will lead to the identity given in equation \WIF.
The expression ${\cal F}_2$ is the correction to the Ward identity.
Simplifying this expression, we obtain
\eqn\WIFltcor{-{\beta\over\alpha}\left(1+ ({\beta\over\alpha})^2\right)
               k_l F(\vec k - n\hat e_l)
          \left[\epsilon^{\mu_l x}\prod_{i=1\atop i\ne l}^N r^{\mu_i x}(k_i)
            +\epsilon^{\mu_l y}\prod_{i=1\atop i\ne l}^N r^{\mu_i y}(k_i)
           \right].}
This cancels with the correction from the second term in the Ward identity
\WIFstcor.  Therefore, even when $0< k_l < n$, the Ward identity for
$F$, given by equation \WIF, still holds.

When $N=2$, there are additional corrections to the Ward identity, because
now the last term also contains two-point functions, and
the second term contains a product of two two-point functions.
 Once again, it is
straightforward to show that these new corrections still cancel each other.
Therefore, for any $N$ and $\vec k$, the Ward identity is given by equation
\WIF.

Equation \WIF\ is important because it gives a relation between $(N+2)$-point
functions and functions with fewer variables.  Unfortunately, as mentioned
earlier, the identities
for different values of $n$ are not independent, so they do not give enough
information to solve for the correlation functions.  Instead, all the
information is contained in the $SL(2,R)$ equations and the Ward identity
with $n=2$, which has the form
\eqn\WInt{F(\vec k, -1, -1; \beta)
         = {1\over 2}  \sum_{s\subset S} F(\vec k_s, -1;\beta)
               F(\vec k_{\bar s}, -1; \beta)
          -{1\over 2} \sum_{i=1}^N k_i F(\vec k -2\hat e_i;\beta).}
Another important feature of equations \WIF\ and \WInt\ is that their form
does not depend on $\beta$.  This property will insure that if the
symmetries do determine the correlation functions, then the exact form of
the duality transformation is satisfied, up to one overall renormalization
constant.


\newsec{Consequences of SL(2,R) Invariance}
When correlation functions are SL(2,R) covariant, it is well known that we
can use the symmetry to fix three of the coordinates.  This uniquely determines
the 2 and 3-point functions up to normalizations, and the 4-point function
up to an arbitrary function of the cross-ratio of the coordinates.

Once we work in momentum space, it is no longer so clear what the SL(2,R)
symmetry implies.  In this section we will show how the SL(2,R) symmetry,
given by
\eqn\sltrm{\sum_{i=1}^N k_i F(\vec k - \hat e_i)=0 \quad {\rm with}
     \quad  \sum_{i=1}^N k_i =  1,}
and
\eqn\sltrp{\sum_{i=1}^N k_i F(\vec k + \hat e_i)=0 \quad {\rm with}
     \quad  \sum_{i=1}^N k_i = -1,}
restricts the possible forms of the correlation functions.  In the following
calculations, we will also assume that the symmetries and boundary conditions
of Section 3.3 hold.  The results for the general case are only a slight
modification of the ones given in this section.

\subsec{Exactly One Negative $k_i$}
First we will show that $F(\vec q) = 0$ if one component of $\vec q$ is
negative and all others are positive.  Let $q_1$ be the component that is
less than zero.  Then we have $q_i \ge 0$ for $i\ne1$.
We will induct on $P$, the absolute value of the negative component of
$\vec q$.

When $P = |q_1| = 1$, by translational invariance the only possibility for
$\vec q$ is $\vec q = (-1,1, 0,\dots, 0)$, up to permutations of the $q_i$.
In this case, $F(\vec q) = 0$ as long as $N>2$.  When $N=2$, instead
$F(\vec q) \ne 0$, which is what we expect because it  gives us the
two-point function.  When $q_1 = -2$,
the two possibilities for $\vec q$ are $\vec q = (-2,2,0,\dots, 0)$ and
(-2,1,1,0,\dots,0).  For $N\ge 4$, each of these have a component equal to
zero, so again $F(\vec q) = 0$.

Now suppose $F(\vec q) = 0$ for any vector $\vec q$ satisfying
$|q_1| < P$, for some $P > 1$.  We will
consider $F(\vec p)$ where the first component of $\vec p$ is $p_1 = -P$,
and we will define $\vec k$ by $\vec k = \vec p + \hat e_1$.  Then we can
write equation \sltrm\ as
\eqn\FpoFpi{F(\vec p) =-{1\over k_1} \sum_{i=2}^N k_i F(\vec p(i)),}
where $\vec p(j) = \vec k - \hat e_j$.  The first component of each
$\vec p(i)$ is given by $p_1(i) = k_1$.  Using the definition of $\vec k$,
we find $p_1(i) = p_1 +1$. Because $p_1$ is negative, this implies that
the absolute value of the negative component of each $\vec p(i)$ on the
right-hand side of equation \FpoFpi\
is given by $P-1$.  It follows that, by the induction hypothesis,
$F(\vec p(i))=0$.  Therefore, the SL(2,R) invariance implies that if one of
the $k_i < 0$ and all the others are greater than zero, or {\it vice versa},
then $F(\vec k) = 0$.

\subsec{General $\vec k$; $F(1, k_2, \dots, k_{N-1}, -1)$}
Next we will consider the remaining case, when at least two $q_i$ are
positive and at least two $q_i$ are negative.  We will show that in that
case $F(\vec q)$ for $\sum q_i = 0$ can always be written as a sum
of $F(\vec k(j))$'s, where each $\vec k(j)$ has one component equal to one and
another component equal to minus one.

Suppose $q_1$ is the smallest positive
$q_i$.  We will first induct on this smallest component and show that
$F(\vec q)$ can be written as a sum of $F(\vec q(j))$'s where each
$\vec q(j)$ has its first component equal to one.  When $q_i = 1$, this is
automatically satisfied.

Now suppose there is some $P>1$ such that, for any $q_1 < P$,
we can write $F(\vec q)$
as a sum of $F(\vec q(i))$'s where the first component of each
$\vec q(i)$ is equal
to one.  Consider $F(\vec p)$, with $p_1$ equal to $P$.  We will define
$\vec k$ by
\eqn\kqme{\vec k = \vec p - \hat e_1.}
Then we can write equation \sltrp\ as
\eqn\FqoFpi{F(\vec p) = -{1\over k_1} \sum_{i=2}^N k_i F(\vec p(i)),}
where
\eqn\pjkej{\vec p(j) = \vec k + \hat e_j.}
On the right-hand sideof equation \FqoFpi,
the first component of each $\vec p(i)$ is given by
$p_1(i) = k_1$.  According to our definition of $k_1$,
we find that $p_1(i) = p_1-1 = P-1$. By our induction hypothesis, we can
then write each $F(\vec p(i))$ as a sum of $F(\vec q(i))$'s where each
$q_1(i)$ is equal to one.  Therefore, the same is true for $F(\vec p)$.
Note that in equation \FqoFpi, for each $\vec p(i)$ we have $p_i(i) > p_i$.
This implies that we cannot reduce any other component of $\vec p$ to
one while keeping $p_1$ equal to one.

Instead, we will next show that any $F(\vec q)$ where $\vec q = (1, q_2,
\dots, q_N)$ can be written as a sum of $F(\vec q(i))$ where each
$\vec q(i)$ has one component equal to one and the other equal to minus
one.  We will now suppose that $q_N$ is the negative component of $\vec q$
with the smallest absolute value.  If $q_N = -1$, we are done.  For the
general case, we will suppose that our hypothesis is true for any $\vec q$
with $|q_N| < P$, for some $P>1$.  Then we will consider $F(\vec q)$ with
$|q_N| = P$.  This time we will define $\vec k$ and
$\vec p(i)$ by $\vec k = \vec q + \hat e_N$ and
$\vec p(i) = \vec k - \hat e_i$.  According to equation \sltrm, $F(\vec q)$
satisfies
\eqn\FqNFpi{F(\vec q) = -{1\over k_N} \sum_{i=1}^{N-1} k_i F(\vec p(i)).}
Because $q_1=1$, the first component of each $\vec p(i)$ is $p_1(i) = 1$
for $i \ne 1$, and $p_1(1) = 0$.  Thus, in the first case, $p_1(i)$ remains
equal to one, as desired, and in the second case $F(\vec p(1)) = 0$,
so it drops out.  The last component of each $\vec p(i)$ is given by
$p_N(i) = q_N+1$.  This implies that $|p_N(i)| = P-1$, which in turn
implies that, by our induction hypothesis, each $F(\vec p(i))$ can be written
as a sum of $F(\vec q(i))$'s where $q_1 =1$ and $q_N = -1$.

Therefore, SL(2,R) invariance tells us that if at least two $k_i$ are
positive and at least two are negative, we can always get one $k_i$ equal
to one and another $k_i$ equal to minus one.  In other words, all non-zero
$F(\vec k)$ can be written in terms of $F(1, k_2, \dots, k_{N-1}, -1)$
for arbitrary $(k_2, \dots, k_{N-1})$.

\subsec{General $\vec k$; $F(a,a,k_3, \dots, k_{N-2}, b, b)$}
The form we just found for $F$ is not the unique way to fix $F$ in momentum
space.  Often it is more convenient to express $F(\vec k)$ in terms of
$\vec k$ which have the form $\vec k = (a, a, k_3, \dots, k_{N-2}, b,b)$,
where $k_1 = k_2 = a$ are the two largest positive components of $\vec k$,
and $k_{N-1} = k_{N} = b$ are the two negative components of $\vec k$
with the largest absolute value.  We will show this is true using a
proof similar to the previous two.

We will first show that $F(\vec q)$ can
always be written as a sum of $F(\vec q(i))$ where the two largest
positive components of $\vec q(i)$ are equal.  To show this,
we will induct on the difference between the two largest positive
components of $\vec q$.  We will assume that the largest component is $q_1$
and the second largest is $q_2$.  This can always be arranged by the
permutation symmetry of $F(\vec k)$.  If $q_1 = q_2$, we are done.  Next,
we consider $q_1 = q_2 + 1$.  We will define $\vec k$ as in equation
\kqme.  In this case, $\vec k$ can be written as $\vec k = (k_1, k_1, k_3,
\dots, k_N)$.  Then equation \sltrp\ says that
\eqn\FqFkFk{k_1 F(\vec q) = -k_1 F(k_1, k_1+1, k_3, \dots, k_N)
                        - \sum_{i=3}^N k_i F(\vec k+ \hat e_i).}
Because $F$ is symmetric under interchange of the $k_i$, this implies
\eqn\FqtFpi{k_1 F(\vec q) = -{1\over 2} \sum_{i=3}^N k_i F(\vec p(i)),}
where $\vec p(i) = \vec k + \hat e_i$ and $p_1(i) = p_2(i) = k_1$.
Therefore, our claim is true in this case.

Now suppose that there is a $P> 1$ such that, for any vector $\vec q$
with $q_1 - q_2 < P$,
we can write $F(\vec q)$ as a sum of $F(\vec q(i))$'s where the two
largest components of each $F(\vec q(i))$ are equal.  We can consider
$F(\vec p)$ with $p_1 - p_2 = P$ and define $\vec k$ and $\vec p(i)$ as in
equations \kqme\ and \pjkej.  When we apply equation \FqoFpi\ to
$F(\vec p)$, on the right-hand side the first component of each $\vec p(i)$
will equal $p_1-1$, and all the other components of $p_i$ will remain
constant or increase by one.  Therefore, for each $i$, we have
$p_i(1) - p_i(2) < P$.  By the induction hypothesis, $F(\vec p(i))$ can then
be written as a sum of $F(\vec q(i))$'s with the two largest components of
each $\vec q(i)$ equal.  Thus, the same is true of $F(\vec q)$.  (Similar
calculations show it is impossible to decrease $q_1$ and $q_2$ any further
while keeping them equal to each other.)

Because $F(\vec q) = F(-\vec q)$, we can also write $F(\vec q)$ as a sum of
$F(\vec q(i))$'s where the two negative components with largest absolute
value are equal.  Now all that remains is to show that we can simultaneously
set the two largest positive $q_i$ equal and the two largest negative $q_i$
equal.  To show this, we will induct on $S$, given by
\eqn\Sabsq{S = \sum_{i=1}^N|q_i|.}
Suppose $\vec q$ has $L$ positive components and $M$ negative components
with $L+M=N$ and $L \ge M$.  Then the smallest possible value of $S$ that
will give a non-zero $F(\vec q)$ is $S = 2L$.  For this value of $S$,
the vector $\vec q$ has the form
$\vec q = (1, \dots, 1, q_{L+1}, \dots, q_N)$.  We can always write
this $F(\vec q)$ as as sum of $F(\vec q(i))$'s where the largest two
negative components of $\vec q(i)$, $q_{N-1}(i)$, and $q_N(i)$ are equal,
and $q_j(i) =1$ for $1 \le j \le L$.  This is because, every time we apply
equation \FqNFpi\ to reduce $|q_N|$ by one, the components of $\vec q$
that are one either remain one or become zero.

Next, we will consider $F(\vec q)$ where $\vec q = (q_1, q_2, \dots,
q_{N-2}, q_N, q_N)$ with $|q_N| > |q_i|$ for $1 \le i \le N-2$.  Suppose
all such $F(\vec q)$ with $\sum_{i=1}^N |q_i| < S$ for some $S$ can be
written in the desired form.  Then take $\vec q =(q_1, q_2, \dots, q_N, q_N)$
with $\sum_{i=1}^N |q_i| = S$.  We can keep applying equation \sltrp\
or \sltrm\ to $F(\vec q)$ until it is written as a sum of $F(\vec p(i))$'s
with $p_1(i) = p_2(i)$.  At this point, some of the $\vec p(i)$'s
will no longer have $p_{N-1}(i) = p_N(i)$.
However, for each such $\vec p(i)$, the sum of the
absolute value of its components is smaller than the
original value of $S$ for $\vec q$.  Therefore, by our induction hypothesis,
we are done.

We conclude that SL(2,R) invariance lets us set the two largest positive
$k_i$ equal and the two largest negative $k_i$ equal.

\subsec{Continuity of $F(\vec k)$}
The SL(2,R) symmetry combined with the homogeneity and piecewise linearity
(of the form in equations \FaRk, \sumk, and \sumkzero)
of $F(\vec k)$ also requires $F(\vec k)$ to be a continuous function when
the $k_i$ take on real values.  In this section we will show how the
SL(2,R) invariance implies that when $\vec k$ crosses a simple boundary
between a region where $F$ has one slope and a region where $F$ has another
slope, $F(\vec k)$ is continuous.  We will not consider what happens
when $\vec k$ lies in such a boundary and crosses intersections of
these boundaries, since, although the basic idea is the same, it is much
more complicated, and, also, the calculations of $F(\vec k)$ in reference
\contact\ can be used to directly show that it is continuous.

According to equations \FaRk, \sumk, and \sumkzero\ for $F(\vec k)$, the
slope of $F(\vec k)$ jumps whenever $\sum_{i\in S} k_i = 0$, for
any $S \subset \{1, \dots, N\}$.
We will consider the particular boundary $B$, given by
$\sum_{i\in S} k_i = 0$ for some particular $S$ that contains $1$ and does
not contain $N$.  The two regions, $R_+$ and $R_-$, on either side of the
boundary have
\eqn\sumkRp{\sum_{i\in S} k_i > 0 \quad {\rm for \quad all} \quad
\vec k \in R_+,}
and
\eqn\sumkRm{\sum_{i\in S} k_i < 0 \quad {\rm for \quad all} \quad
\vec k \in R_-.}
Let $\vec v$ be a vector that lies in the boundary.  We can assume that all
the partial sums other than $\sum_{i\in S} k_i$ are either greater than one
or less than one.  This is because if $\vec v$ is not in the intersection
of boundaries, then only the one partial sum (and its complement)
can equal zero.  If any of the other partial sums equal $1$ (for $k_i \in Z$),
we can just multiply $\vec v$ by a constant and use the resulting vector
instead.  (This condition is necessary to guarantee that all the $\vec p(i)$
defined below lie only in $B$ or $R_+$.)
We will define $\vec k$ by $\vec k = \vec v + \hat e_1$, and $\vec p(i)$
by $\vec p(i) = \vec k - \hat e_i$.  Then for $i\in S$, the vector
$\vec p(i)$ lies in the boundary because $\sum_{j\in S} p_j(i) = 0$.
For $i \notin S$, the vector $\vec p(i)$ has $\sum_{j \in S} p_j(i) = 1$,
so it lies in the region $R_+$.

The SL(2,R) equation for $F(\vec v)$ is
\eqn\sltrFv{(v_1 + 1) F(\vec v)
     + \sum_{i \in S \atop i \ne 1} v_i F(\vec k - \hat e_i)
     + \sum_{i \notin S} v_i F(\vec k - \hat e_i) = 0.}
Now we can use the piecewise linearity of $F(\vec k)$.  We have
$F(\vec k) = \vec a_R \cdot \vec k$, where $\vec a_R$ depends only on the
sign of the partial sums.  Therefore, we can take $F(\vec k) = \vec a \cdot
\vec k$ in the region $R_+$, and $F(\vec k) = \vec b \cdot \vec k$ in the
boundary.  Keeping track of the different regions, we can write the SL(2,R)
equation as
\eqn\sltrabv{(v_1 + 1) \vec b \cdot \vec v
  + \sum_{i \in S \atop i \ne 1} v_i \vec b \cdot (\vec v+\hat e_1-\hat e_i)
     + \sum_{i \notin S} v_i \vec a \cdot (\vec v + \hat e_1 - \hat e_i)
      = 0.}
If we expand this equation out and use the relation
$\sum_{i\in S} v_i = \sum_{i\notin S} v_i = 0$, we find that
\eqn\abvcalc{\vec b \cdot \vec v - \sum_{i \in S} v_i b_i
                 - \sum_{i\notin S} v_i a_i = 0.}
We can combine the sum and dot product to obtain
\eqn\vbvaS{\sum_{i \notin S} v_i b_i = \sum_{i\notin S} v_i a_i.}

Now we can repeat the same calculation, but with $\vec k = (v_1, \dots,
v_N-1)$, where $N\notin S$.  This time we will define $\vec p(i)$ by
$\vec p(i) = \vec k + \hat e_i$.  The components of $\vec p(i)$ now
satisfy $\sum_{j\in S} p_j(i) = 1$ for $i\in S$ and
$\sum_{j\in S} p_j(i) = 0$ for $i \notin S$.  This means that $\vec p(i)$
is in $R_+$ for $i\in S$, and $\vec p(i)$ is in the boundary for
$i \notin S$.  If we replace $S$ by its
complement $\bar S$, instead we have $\vec p(i) \in B$ for $i \in \bar S$
and $\vec p(i) \in R_+$ for $i \notin \bar S$.  Thus the roles of $S$ and
$\bar S$ are reversed from the first calculation.  In this new calculation,
we must use the other SL(2,R) equation for $F(\vec v)$.  However, because
$F(\vec v) = F(-\vec v)$, equation \sltrp\ is equivalent to equation
\sltrm\ if we just replace $\vec v$ with minus $\vec v$.  It follows that if
we repeat the calculation with this new $\vec k$, we will obtain the same
result as in equation \vbvaS, except that $S$ is replaced with its complement
and $\vec v$ is replaced with $-\vec v$.  Thus, we have
\eqn\vbvaSb{\sum_{i \in S} v_i b_i = \sum_{i\in S} a_i v_i.}
Adding equation \vbvaS\ and \vbvaSb, we obtain
\eqn\vbva{\vec b \cdot \vec v = \vec a \cdot \vec v.}
For every vector $\vec v$ in the boundary, this equation is satisfied
by either $\vec v$ or some multiple of $\vec v$.  Because in each region $F$
is a linear function, equation \vbva\ says that if we let the $v_i$ take
on values in the reals, as $\vec v$ goes from $R_+$ to $B$ the limit of
$F(\vec v)$ defined on $R_+$ equals $F(\vec v)$ defined on $B$.  A similar
calculation shows the same is true for $R_-$ and $B$.  Therefore, $F$ is
continuous when crossing the hyperplane $B$.

The proof that $F$ is continuous when $\vec v$ crosses from a boundary
region (or intersection of boundary regions) to another boundary region
(or intersection of boundary regions) uses the SL(2,R) invariance and
linearity in the same way, but it is much more complicated to keep
track of the regions.

\newsec{Exact Solutions for Correlation Functions}
In this section, we will use all the properties of $F(\vec k)$ derived in
the previous sections to solve for the correlation functions.  These
properties will give us the solution for the two-point function up to
normalization, $\mu$, and they determine the four and six-point functions
in terms of $\mu$.  We can also solve for all $F(\vec k)$ where two $k_i$
are positive and all others are negative, and {\it vice versa}.

\subsec{SL(2,R) Functions Satisfying the Boundary Conditions}
In this section we will begin by showing that the simplest expression
that is continuous, piecewise-linear, SL(2,R) covariant, and that satisfies
the boundary conditions given in Section 3.3, has the form
\eqn\Pkdef{P(\vec k) = \sum_{s\subset S} (-1)^{|s|+1}
\left|\sum_{i\in s} k_i \right|,}
where $S = \{1,2, \dots,N\}$ and $|s|$ is the number of elements in $s$.
First, we note that $F(\vec v) = \vec a \cdot \vec v$ and
$F(\vec v) = \left | \sum_{i\in s}  v_i \right| $ both satisfy the SL(2,R)
equations.  This can be verified by substitution into equation \sltrm\
or \sltrFv.  In addition, when we set one $k_i$ to zero, we find $P(\vec k)
=0$, so the boundary conditions are also satisfied.  Therefore $P(\vec k)$
satisfies all the requirements on the correlation functions that we
derived in the previous sections, except possibly the $n = 2$ Ward identity.
The correlation functions given in references \contact, \cgcirk, and
\ckmy\ all have
this form.  However, in the more general case, this is not the only
continuous, piecewise-linear function with the appropriate boundary
conditions that satisfy the SL(2,R) equations.

To get an idea of what the other solutions are like, we will analyze
$P(\vec k)$ further.  It has the form of a ``symmetrization" of
the SL(2,R) invariant two-point function, $P(k,-k) = 2|k|$, over
the $N$ components of $\vec k$.  Also, if we multiply $P(\vec k)$ by
products of $\sign(k_i)$, it is still continuous and satisfies the
boundary conditions.  The only correlation functions obtained from
$P(\vec k)$ that are not just contact terms contain exactly two $x$'s or
exactly two $y$'s.  Such correlation functions go as
\eqn\Corrxx{\langle y(t_1) y(t_2) x(t_3) \dots x(t_N) \rangle
=  c {1\over (t_1 - t_2)^2}\prod_{j=3}^N
   \left({1\over t_j - t_1} - {1 \over t_j - t_2}\right),}
where $c$ is a constant, independent of the $t_j$.  This has the same form
as the {\it free} correlation function $\langle \cos(x(t_1)) \cos(x(t_2))
\dot x(t_3) \dots \dot x(t_N) \rangle$.

These remarks suggests what happens in the general case.  First, if we take
any continuous, piecewise-linear, SL(2,R) covariant function of $M$ variables
that vanishes when
any one of the variables equals zero, and then ``symmetrize" this function
over $N$ variables, the resulting function will still have all of the
desired properties. Similarly, any free correlation function of the form
$\langle \prod_{j=1}^{M} \cos(x(t_j))
\prod_{j=M+1}^N \dot x(t_j) \rangle$ is always SL(2,R) invariant and has
a piecewise-linear Fourier transform.  It is possible to add in some contact
terms so that this Fourier transform also satisfies the boundary conditions.
The questions that remain are whether the correlation functions that satisfy
all the Ward identities are always made up of functions of the form
described above, and to what extent the correlation functions are determined
exactly by all the symmetry conditions derived in this paper.  In the
remaining  part of this paper, we will address these issues by using the
symmetries to derive various correlation functions.  Since the work in this
paper was done, the authors of \ckmy\ have shown in that paper
that the correlation functions of $\dot x(t)$ and $\dot y(t)$
must indeed be given by free correlation functions of $e^{\pm ix(t)}$ and
$\dot x(t)$.

\subsec{Two-point Function}
For the two-point function, the only homogeneous, piecewise-linear
solution for $F$ that is even in $\vec k$ and is translation invariant
is
\eqn\Ftp{F(\vec k) = \mu |k|.}
In this equation, $k = k_1 = -k_2$ and  $\mu$ is an arbitrary constant that
depends on $V_0$ and possibly on how the theory is renormalized.
In dissipative quantum mechanics, $\mu$ plays the role of the mobility of
the particle \schmid.  The methods of calculation in this paper do not
give us the value of $\mu$ in terms of $V_0$.  However, because the
two-point function has the form given in \Ftp\ (with non-zero $\mu$),
the particle is delocalized at the special multicritical points

This solution for $F$ in equation \Ftp\
also follows directly from SL(2,R) invariance combined with the symmetry
under $\vec k \to - \vec k$.  If we substitute $F(\vec k)$ for
$\tilde C(k, -k;0)$ in equation \Ctptrans\ and simplify the resulting
expression, then we find that the two-point function is given by
\eqn\tCtp{\tilde C^{\mu\nu}(k, -k; \beta) =
 \delta^{\mu\nu}\left[\left(1-({\beta\over\alpha})^2\right)\mu
   + ({\beta\over\alpha})^2 \right] |k|
  + \epsilon^{\mu\nu}{\beta\over\alpha} \left(2\mu - 1\right) k.}
When $V_0=0$, $\tilde C^{\mu\nu}$ must reduce to its value for the
free theory, $\tilde C^{\mu\nu} = \delta^{\mu\nu}|k|
+ \epsilon^{\mu\nu}(\beta/\alpha)k$.  This implies that when $V_0 = 0$
(and also to order $V_0^0$ in perturbation theory) we have $\mu = 1$.

When we take the Fourier transform of this function for $\tilde C$, we find
that in real space, as $T \to \infty$, the correlation function is given
by
\eqn\Ctpt{C^{\mu\nu}(t_1-t_2) =
2\mu_\alpha{1\over (t_1-t_2)^2} \delta^{\mu\nu} -
    \epsilon^{\mu\nu} i\pi\mu_\beta \delta'(t_1-t_2) ,}
where $\mu_\alpha = \left(1-(\beta/\alpha)^2\right)\mu + (\beta/\alpha)^2$
and $\mu_\beta = 2(\beta/\alpha)(2\mu-1)$.
The first term is what we expect from the SL(2,R) symmetry, but the second
term clearly also transforms properly under SL(2,R) transformations.
Lastly, because the two-point function is fixed by the
SL(2,R) invariance, equation \Ctptrans\ guarantees that the duality
transformation will also be satisfied.

\subsec{Four-point Function}
Next, we calculate the four-point function.  There are two different types
of regions for $\vec k$ we must consider.  In each of these regions,
it is straightforward to solve for the continuous,
piecewise-linear functions, up to normalizations, given the boundary
conditions $F(\vec k) = 0$ when $k_i = 0$.  Then the $n=2$ Ward identity
fixes the normalization.  The solution for $F$ is given by
\eqn\Ffpsum{F(k_1, k_2, k_3, k_4) = \delta_{k_1+k_2+k_3+k_4}
     {1\over 4}(\mu^2 - \mu)
    \sum_{s \subset \{1,2,3,4\}} (-1)^{|s|+1} \left|\sum_{i\in s}k_i\right|,}
where $|s|$ is the number of elements in $s$.  Thus $F$ is proportional to
$P(\vec k)$, given by equation \Pkdef.
This expression for $F$ is equivalent to
\eqn\Ffpmin{\eqalign{F(k_1,k_2,k_3,k_4) =&  \delta_{k_1+k_2+k_3+k_4}
     (\mu^2 - \mu) {\rm min}\left(|k_1|,|k_2|,|k_3|,|k_4|\right) \cr
 &\times {1\over 2}\left(1 + \sign(k_1)\sign(k_2)\sign(k_3)\sign(k_4)\right).}}

Alternatively, the SL(2,R) invariance, the homogeneity of $F$, and the
$n=2$ Ward identity are enough to uniquely determine $F$ for the four-point
function.  This can be shown as follows.  The first of the two types of
regions for $\vec k$ has $k_1k_2k_3k_4 < 0$, which can only happen if
either exactly one $k_i$ is positive or exactly one $k_i$ is negative.
In both cases, the SL(2,R) symmetry requires $F(\vec k) = 0$.  The  second
type of region has $k_1k_2k_3k_4 > 0$, which means two $k_i$ are positive
and two are negative.  For concreteness, suppose $k_1$, $k_2 > 0$ and
$k_3$, $k_4 <0$.  We can use the SL(2,R) symmetry to set $k_1=k_2 =a$
and $k_3 = k_4 = b$ for some $a > 0$ and $b < 0$.  Momentum conservation
then implies that $a= -b$.  Thus all $F(\vec k)$ are determined once
$F(a,a, -a,-a)$ for all positive integers $a$ is known.  This function
satisfies
\eqn\FaaFo{F(a,a,-a,-a) = a F(1,1,-1,-1)}
because $F$ is homogeneous in $\vec k$.  Therefore, the only unknown is
$F(1,1,-1,-1)$ and permutations.  We can solve for $F(1,1,-1,-1)$
in terms of the two-point function by using the $n=2$
Ward identity.  If we set $\vec k  = (1,1)$ and $N = 2$, then equation
\WInt\ becomes
\eqn\FfpFtp{F(1,1,-1,-1) = F^2(1,-1)
   - {1\over2} \left(F(-1,1) + F(1,-1)\right).}
Substituting in $F(1,-1) = \mu$, we obtain
\eqn\Ffpone{F(1,1,-1,-1) = \mu^2 - \mu.}
Therefore, the four-point function is {\it uniquely} determined given the
two-point function.  Once we have this solution for $F(1,1,-1,-1)$, we can
``integrate up" the difference equations to obtain the solution \Ffpsum
for $F(\vec k)$.

The solution for $\tilde C$ can be obtained by substituting the expression
for $F$ into equation \CRFdef.
We note that $F$ and $\tilde C$ are identically zero whenever $\mu = 0$ or
$\mu = 1$.  When $\mu = 0$, the two-point function is zero, so all
correlation functions vanish.  The somewhat more interesting case when
$\mu = 1$ just corresponds to the free theory, where we expect the two-point
function to be non-zero, but all other correlations to vanish.

To better understand these solutions, we will transform some special
cases back to real space. When $\beta = 0$, the correlation function is
\eqn\fptbz{\eqalign{\langle \dot x(t_1) \dot x(t_2) \dot x(t_3) \dot x(t_4)
\rangle
   =& {1\over2} (\mu^2 - \mu)\bigg[{1\over t_1-t_2}{1\over t_2-t_3}
   2\pi\delta(t_3-t_4) 2\pi\delta(t_4-t_1)\cr
  & - {1\over t_1-t_2} 2\pi\delta(t_2-t_3)
   {1\over t_3-t_4} 2\pi\delta(t_4-t_1) + {\rm permutations} \bigg],}}
where the only permutations that are included are the ones that
treat points joined by delta-functions as indistinguishable.
{}From this equation, we see that when $\beta = 0$ the correlation function
contains only contact terms.  For non-zero $\beta$ some of the correlation
functions do not contain only contact terms.  To have finite long-time
behavior, $\tilde C$ must depend on three independent variables.  This
occurs only when two of the fields are $x$'s and two of them are $y$'s.
Then, in Fourier space we have
\eqn\tCGF{\eqalign{\tilde C^{xyxy}(\vec k) =& G(k_1,k_2,k_3,k_4)\cr
              = & \left({\beta\over\alpha}\right)^2
                 \left[\sign(k_1)\sign(k_3) + \sign(k_2)\sign(k_4)\right]
                  F(\vec k).}}
In real space, this goes as
$1/\left((t_1-t_2)(t_2-t_3)(t_3-t_4)(t_4-t_1)\right)$.  This is equivalent
to the function given in equation \Corrxx, so it has the same form as
the free correlation function of $\dot x(t_1)$ and $\dot x(t_3)$ with
$\cos(x(t_2))$ and $\cos(x(t_4))$.

\subsec{Six-point Function and the Self-Dual Point}
For the six-point function, one can again solve for the continuous,
piecewise-linear, homogeneous functions that have the correct boundary
conditions, but it is already quite tedious.  Also,
the SL(2,R) invariance, homogeneity and $n=2$ Ward identity no longer
appear to
uniquely fix the six-point function.  Instead, in this section we will
first use the SL(2,R) invariance to restrict $\vec k$ to only two
types of regions.  Then we will use the linearity and boundary condition
to solve for the function in these regions, and finally we will use the
$n = 2$ Ward identity to set the normalization in each of these regions.

Because of the SL(2,R) invariance, the only choices for the signs of
the $k_i$ that give non-zero $F(\vec k)$ are either three $k_i$'s
positive and three negative, or exactly two $k_i$'s negative.  (By
inversion symmetry, this last case also takes into account the
case when exactly two $k_i$'s are positive.)  The SL(2,R) invariance
further implies that the only unknowns in these regions can be written
as
\eqn\kthree{\vec k_3 = (a,a,b,-c,-d,-d) \quad {\rm with} \quad a \ge b
       \quad  d \ge c,}
and
\eqn\ktwo{\vec k_2 = (a,a,b,c,-d,-d) \quad {\rm with} \quad
a \ge b \ge c,}
where $a$, $b$, $c$, $d > 0$.  Because $F(\vec k)$ is invariant under
inversion, we can assume $d \ge a$ in equation \kthree. These orderings
of the variables $a$, $b$, $c$, and $d$, combined with momentum conservation
$\sum_{i=1}^6 k_i = 0$,
uniquely determine the signs of the partial sums for any $\vec k_2$
or any $\vec k_3$.
Thus the $\vec k_3$'s given in equation \kthree\ all lie in only one region,
and similarly for the $\vec k_2$'s.  When any of the equalities are satisfied
in equation \kthree\ or \ktwo, then $\vec k_3$ or $\vec k_2$,
respectively, lies in the boundary of the region.
Both the region for $\vec k_3$ and the region for $\vec k_2$ also have
the boundary where $c = 0$ (and $a$, $b$, and $d \ne 0$).  This boundary
region contains, for example, the two linearly independent vectors
$(3,3,2,0,-4,-4)$ and $(4,4,2,0,-5,-5)$.

Because of the piecewise linearity of $F$, we know that for
$\vec k_3$ the function $F$ has the form
\eqn\FkvecA{F(\vec k_3) = \vec A \cdot \vec k_3,}
for some constant $\vec A$.  For $\vec k_3$ given by equation \kthree,
the momentum conservation implies
$2a + b - c -2d = 0$. Using this condition
to eliminate $d$ from equation \FkvecA, we find that $F(\vec k_3)$ must
have the form
\eqn\Fknod{F(\vec k_3) = A_a a + A_b b + A_c c = 0,}
for some constants $A_a$, $A_b$, and $A_c$.
The boundary conditions require that $F(\vec k_3) = 0$
whenever $c = 0$.  This implies that
\eqn\Fkboundary{F(\vec k_b) = A_a a + A_b b = 0,}
for any $\vec k_b$ in the boundary.
Since we have at least two linearly independent vectors in the boundary,
equation \Fkboundary\ implies that $A_a =  A_b = 0$.  Therefore, $F(\vec k_3)$
has the form
\eqn\Fkthree{F(\vec k_3) = A_c c.}
The $n = 2$ Ward identity can then be used to fix $A_c$, so that $F(\vec k_3)$
is uniquely determined.
A similar calculation shows that $F(\vec k_2)$ is also uniquely determined.

It follows that in each of these regions, $F$ must be proportional to
$P(\vec k)$.  Using the Ward identity to fix the constants of
proportionality, we find
\eqn\Fsixp{F(\vec k) = -{1\over 8} (\mu^2 - \mu) P(\vec k) \quad {\rm for}
\quad \prod_{i=1}^6 k_i > 0,}
and
\eqn\Fsixm{F(\vec k) = {1\over 8} \mu (2\mu-1)(\mu-1) P(\vec k)
\quad {\rm for} \quad \prod_{i=1}^6 k_i < 0.}
We first note that $F(\vec k)$ is no longer simply proportional to
$P(\vec k)$, because the coefficient now depends on which ``quadrant"
$\vec k$ is in.  However, it does fall under the general form described
in Section 6.1.  When $\beta = 0$ (or if
all the variables are only $x$'s or only $y$'s) the correlation function
in Fourier space, $\tilde C$, depends only on three independent variables,
which implies that in real space it consists of contact terms.
The  only solution without contact terms
is $\langle \dot x(t_1) \dot x(t_2) \dot y(t_3) \dots \dot y(t_6) \rangle$
(and the same with $x$ and $y$ interchanged), and it has the form given in
equation \Corrxx. Thus all the symmetries derived in this paper uniquely
determine the six-point function and require it to have the form described
in Section 6.1.

Lastly, we note that when $\mu = 1/2$, $F(\vec k) = 0$ for $\prod k_i <0$.
Thus, according to equation \sdual\ we expect that the self-dual
point should occur at $\mu = 1/2$.
In fact, using the Ward identity, it is straightforward to show that
$\mu = 1/2$ does behave like the self-dual point; at this value of $\mu$
{\it all} correlation functions with $\prod k_i < 0$ vanish, which means
the $z \to 1/z$ duality transformation is satisfied.  This is the
``self-dual'' value of the mobility found by Schmid in reference \schmid,
and it is rather remarkable that these exact results agree with his
calculations, which involved many approximations.  One
property of the self-dual point that was noted in ref. \cgcdef\
is that at this point the off-diagonal part of the two-point function
vanishes, as can be seen from equation \tCtp.  Because the linear response
to a transverse electric field is related to this correlation function,
these results imply that the ``Hall" current should vanish at the
self-dual points.

\subsec{Exactly 2 $k_i$ are positive}
This time we can restrict the $k_i$ to be $(1, k_2, \dots, k_{n-1}, -1)$,
with $k_2 >0$ and all others negative.  Again, these vectors lie in only
one region and its boundaries.  In particular, they all lie in the
boundary of the region where $k_1$ is less than one
and can be taken to zero without crossing any more
boundaries.  Therefore, as in the case for the six-point function,
$F(\vec k) = a k_1$.  Because $k_1 =1$ in the region we are considering,
$F(\vec k)$ is a constant in this region, and the Ward identity once again
will uniquely determine it.  Thus, we find that
\eqn\Ftkpos{F(\vec k) = (-{1\over 2})^{N/2}(\mu^2 - \mu) P(\vec k),}
whenever exactly two $k_i$ are positive.  This again has the form given
in Section 6.1.

\subsec{Eight-point function and beyond}
For the eight-point function, we find that when exactly four $P(\vec k)$ are
positive, the function $P(\vec k)$ no longer satisfies the Ward identity.
Instead, we must also consider the functions made from appropriately
symmetrizing the four-point function given in equation \tCGF.  This
``symmetrized" function has the form
\eqn\eptextra{\sum_{s_1,s_2,s_3,s_4} (-1)^{|s_1|+ |s_3|}
                G(\sum_{i\in s_1} k_i, \sum_{i\in s_2} k_i,
         \sum_{i\in s_3} k_i, \sum_{i\in s_4} k_i),}
where $s_1$, $s_2$, $s_3$, and $s_4$ are summed over all disjoint
subsets of $S = \{1,2,3,4,5,6,7,8\}$ whose union equals $S$.
Unfortunately, once we reach eight variables, the piecewise-linear functions
become quite complicated, and we cannot easily show that this is the unique
solution.  However, we still conjecture that all correlation functions must
have the form described in Section 6.1, and that the properties and
symmetries derived in this paper are enough to determine all the correlation
functions.  Since this work was originally done, the first of these
conjectures has been shown to be true in reference \ckmy,
but the second still remains an open question.

\newsec{Conclusions}
In this paper, we have used the symmetries of the boundary system with
cosine and magnetic interactions to
find exact solutions for the correlation functions
at the special multicritical points.
We have shown that the piecewise-linearity, homogeneity, reparametrization
invariance Ward identities, and SL(2,Z) duality symmetry uniquely
determine the 2-point, 4-point, and 6-point functions up to normalization,
and also all the correlation functions with exactly one or two positive
momenta.
For the eight-point functions, it again appears likely that these symmetries
determine the correlation functions, and the solution for the eight-point
function suggests the form for all the remaining correlation functions.
Once we get beyond eight variables, the piecewise-linear
functions become rather complicated, so
it is difficult to determine a basis for them.
If the piecewise-linear functions with the required boundary condition
all have the form given in
Section 6.1, it seems likely that the symmetries will continue to uniquely
fix the correlation functions.  In any case, the symmetries combined
with the long-time behavior do appear to fix the contact terms.

The results in this paper also verify that the approximate duality
transformation under $z \to z/(1 + inz)$ is an exact transformation.
The other interesting check on the SL(2,Z) duality symmetry found in
reference \cgcdef\ is what happens under the transformation $z \to 1/z$
when $\beta = 1$.  According to references \schmid\ and \cgcdef,
the value of $V_0$ changes under this transformation, and many
approximations were made in deriving this symmetry.   Remarkably,
we found that the theory at $\beta = 1$ is self-dual precisely
at the value of the mobility predicted in these two
references.  This value of the mobility occurs at a particular value of the
potential strength $V_0$.  We have also found that
the correlation functions at all the other special critical
points with this value of $V_0$ also exactly satisfy the self-dual
condition.   Thus, we expect this value of $V_0$ to play an important role;
one interesting property of the special critical points of the
dissipative Hofstadter model at this potential strength is that
the analogue of the Hall current vanishes.

The consequences of the SL(2,R) invariance and the boundary
reparametrization invariance Ward identities in Fourier space are quite
general, and may be useful for other boundary theories.  In particular,
we note that the equations we used to solve for the correlation functions
have solutions for all values of $\beta/\alpha$, and not just the values
at the special multicritical points.  According to reference \cgcdef,
we expect the dissipative Hofstadter
model to have many other critical and multicritical points at other values
of magnetic flux and friction.
It is interesting to speculate on
whether these other critical theories are described by the additional
conformal theories we have found,
or whether these critical theories have a more complicated structure instead.

\newsec{Acknowledgements}
I would like to thank J.~Cohn and S.~Axelrod for many helpful discussions.

\listrefs
\end